\documentclass[prd, aps, showpacs, preprint, floatfix]{revtex4-1}
\usepackage{hyperref}
\usepackage{epsfig}
\usepackage{graphicx}
\usepackage{amsfonts}
\usepackage{amssymb}
\usepackage{amsbsy}
\usepackage{amsmath}
\usepackage{latexsym}
\usepackage{bbm}
\usepackage{bm}
\usepackage{bbold}
\usepackage{subfigure}
\usepackage{color}
\usepackage{enumitem}

\def\gsim{\raise0.3ex\hbox{$>$\kern-0.75em\raise-1.1ex\hbox{$\sim$}}}
\def\lsim{\raise0.3ex\hbox{$<$\kern-0.75em\raise-1.1ex\hbox{$\sim$}}}

\def\be {\begin{equation}}
\def\ee  {\end{equation}}
\def\bea {\begin{eqnarray}}
\def\eea {\end{eqnarray}}

\begin{document}
\noindent
\title{Polymer Quantization of a Self-Gravitating Thin Shell}

\author{Jonathan Ziprick}
\email{jziprick@unb.ca}
\affiliation{Dept. of Mathematics and Statistics, University of New Brunswick, Fredericton, NB, Canada E3B 5A3}
\author{Jack Gegenberg }
\email{geg@unb.ca}
\affiliation{Dept. of Mathematics and Statistics and Dept. of Physics, University of New Brunswick, Fredericton, NB, Canada E3B 5A3
%\\
% Perimeter Institute for Theoretical Physics, 31 Caroline Street North, Waterloo, Ontario, Canada N2L 2Y5
}
\author{Gabor Kunstatter }
\email{g.kunstatter@uwinnipeg.ca}
\affiliation{ Dept. of Physics and Winnipeg Institute for Theoretical Physics, University of Winnipeg, Winnipeg, Manitoba,Canada R3B 2E9}
%\pacs{04.60.Ds}

\begin{abstract}
We study the quantum mechanics of self-gravitating thin shell collapse by solving the polymerized Wheeler-DeWitt equation. We obtain the energy spectrum and solve the time dependent equation using numerics. In contradistinction to the continuum theory, we are able to consistently quantize the theory for super-Planckian black holes, and find two choices of boundary conditions which conserve energy and probability, as opposed to one in the continuum theory. Another feature unique to the polymer theory is the existence of negative energy stationary states that disappear from the spectrum as the polymer scale goes to zero. In both theories the probability density is positive semidefinite only for the space of positive energy stationary states. Dynamically, we find that an initial Gaussian probability density develops regions of negative probability as the wavepacket approaches $R=0$ and bounces. This implies that the bouncing state is a sum of both positive and negative eigenstates.
\end{abstract}
\date{\today}

\maketitle
%\tableofcontents
\section{Introduction}
The quantum mechanical regime of gravitation continues to be opaque to our probing, both theoretically and experimentally, in the latter case due to the extreme weakness of the gravitational interaction.
 There are now many theoretical probes, including (but not limited to) string theory, loop quantum gravity, spin foams, causal sets, causal dynamical triangulations and asymptotic safety. However, none of these are yet complete or able to predict phenomena which can be seen experimentally. Given the lack of experimental direction, we must rely heavily on philosophy and mathematics to justify any particular approach.  These circumstances have motivated the careful study of mathematical toy models, often obtained by imposing symmetry and/or changing the number of dimensions of spacetime.  Such models, if exactly solvable, could help us to see around the obstacles occluding our view of the quantum regime of realistic theories of gravity.   It is also possible that results obtained in the toy models might survive the transition to realistic models, and provide the basis for attempts to experimentally test quantum gravity.

Gravitational collapse of a spherically symmetric thin shell of dust in 3+1 dimensions presents a simple model of a matter--gravity interaction that has both classical and quantum solutions. Classically, the theory is parametrized completely by a pair of phase space variables: the radius $R$ of the shell and its conjugate momentum $P$. This is effectively a theory of particle mechanics, yet it is rich enough to describe dynamical black hole formation, including the evolution of apparent horizons and singularity formation. On the quantum side, this is a rare solvable model which describe a fully quantum interaction between gravity and matter. As such significant attention has been given to the quantum theory\cite{Hajicek, HKK, Berezin, Stojkovic, Vachaspati, Xu}. Perhaps the most influential work in this area is that of H\'aj\'i\v{c}ek, Kay and Kucha\v{r} \cite{HKK} where a Wheeler-DeWitt quantization was successfully completed for shells with a rest mass $m<1$ (in Planck units). Using a particular choice of time coordinate $t$ and a separation ansatz $\Psi(t,R) = e^{iEt} \psi(R)$, the authors were able to solve for the complete spectrum of time-independent bound and scattering states.

In the present paper, we study quantization of the thin shell model along the same line as H\'aj\'i\v{c}ek \textit{et al}. The key difference is that we employ polymer quantization \cite{poly} rather than standard techniques which assume that spacetime is a continuum. Polymer quantization posits that space is fundamentally discrete near or below some microscopic scale $\mu$, and that the continuum emerges only as a coarse-grained approximation. Our results agree with H\'aj\'i\v{c}ek \textit{et al} in the $\mu \to 0$ limit, and we find interesting new results for the case of small but finite $\mu$. In particular we are able to consistently quantize the theory for rest mass $m>1$, and using numerical methods we are also able to solve the fully quantum equations of motion.

\section{Classical theory}
In this section we establish the classical theory upon which the quantum theory is built. We begin by sketching a derivation of the classical equations of motion for a thin shell of dust collapsing under its own gravitation. Details can be found, for example, in chapter three of Eric Poisson's text book \cite{eric}. After these equations of motion are obtained we develop the canonical theory to set up quantization in the next section.

Now consider the spacetime corresponding to a collapsing thin shell, which represents a codimension-one hypersurface of spacetime. Outside of the shell the spacetime is described by the Schwarzschild metric, while inside the shell spacetime is flat. The `in', 'out' and `shell' line elements are:
\bea
ds^2_{\mbox{\tiny in}} &=& - dt^2 + dr^2 + r^2 d \Omega^2 \\
ds^2_{\mbox{\tiny out}} &=& - f(r) dT^2 + f^{-1}(r) dr^2 + r^2 d \Omega^2 \\
ds_{\mbox{\tiny shell}}^2 &=& - d \tau^2 + R^2 d \Omega^2
\eea
where $f(r) \equiv 1-2E/r$ and $E$ is the \textit{total} gravitational energy of the shell.

The next step is to impose the junction conditions at the shell in order to form a complete spacetime. These conditions require that the metric induced on the shell hypersurface $h_{ab}$ from the `out' spacetime agrees with the metric induced at the shell from the `in' spacetime. Furthermore, the junction conditions tell us that a non-trivial stress-energy is associated with the shell, given by
\be
\label{S1}
T_{ab} = \frac{1}{8 \pi}\left([K_{ab}]-[K]h_{ab} \right),
\ee
where square brackets denote the difference between the outer and inner values of the extrinsic curvature $K_{ab}$ and its trace $K$.

We take the stress energy of the shell to be that of a perfect fluid with 2D density $\rho$ travelling with the velocity $u^a$ of the shell
\be
\label{S2}
T_{ab} = \rho u_a u_b.
\ee
Equating the stress energy given in (\ref{S1}) with that of (\ref{S2}) is most easily done in terms of `shell' coordinates ($R, \tau$). Doing so reveals that: 1) the rest mass $m:=4 \pi R^2 \rho$ of the shell is a constant of motion; 2) the first integral of the radial equation of motion is
\be
E= m\sqrt{1+\left(\frac{d R}{d \tau}\right)^2}-\frac{m^2}{2R}=\hbox{constant}.
\label{eq:M1}
\ee

The velocity in the above equation is determined using the time coordinate $\tau$ of a co-moving observer on the shell. This implies a foliation of the spacetime into hypersurfaces labelled by constant values of $\tau$. Going to a canonical theory in terms of $R$ and its conjugate momentum $P$, one finds that the Hamiltonian is not easily represented as an operator in the quantum theory. Following the steps outlined below, one arrives at the following Hamiltonian:
\be
H = m \cosh\frac{P}{m} - \frac{m^2}{2R}.
\ee
In the quantum theory, the $\cosh$ term must be treated as an infinite sum which is difficult to work with, although significant progress was made by Hajicek\cite{Hajicek}. If we instead choose the inner flat space time coordinate $t$, we obtain a Hamiltonian that  is more manageable.

The relationship between the two time coordinates in question is $d \tau^2 = -dt^2+dR^2$, from which one can derive
\be
\left(\frac{dR}{d\tau}\right)^2 = \frac{\left(\frac{dR}{dt}\right)^2}{\left( 1- \left(\frac{dR}{dt}\right)^2\right)}.
\ee
Putting this into (\ref{eq:M1}) gives:
\be
\label{basic}
E = \frac{m}{ \sqrt{1-\dot{R}^2}}- \frac{m^2}{2R}.
\ee
where from now on a dot represents a derivative with respect to the inner time coordinate $t$. This is the energy of a relativistic particle in a Coloumb potential $V = \frac{m^2}{2R}$.

In order to arrive at the canonical theory, we take the total energy to be the Hamiltonian, that is, $E \to H$, and find the momentum which is conjugate to $R$ for this Hamiltonian:
\be
P = \int \frac{dH}{\dot{R}} = \sqrt{(H+V)^2-m^2}.
\ee
From this we obtain the following expression:
\be
\label{Htilde}
\left(H + V\right)^2 = P^2+m^2 .
\ee

At this point we could take the phase space parametrized by $(R,P)$, solve for the Hamiltonian $H$ in terms of these variables and proceed with quantization.
However, the resulting Hamiltonian has a square root which is difficult to handle in the quantum theory. To avoid this problem we proceed by putting the Hamiltonian into parametrized form, as in \cite{HKK}, by extending the phase space to include the time $t$ and its momentum $p$. The equation (\ref{Htilde}) is treated as a constraint as described in the remainder of this section.

The extended phase space has two canonical pairs $(R,P)$ and $(t,p)$ where the momentum conjugate to time is $p = -H$. The canonical action is
\be
I = \frac{dR}{ds}P + \frac{dt}{ds}p - \lambda h,
\ee
where $s$ is the evolution parameter and $\lambda$ is a Lagrange multiplier. The dynamics is generated by the Hamiltonian constraint
\be
\label{sh}
h:= -(p-V)^2+P^2+m^2,
\ee
which is quadratic in the momenta and does not have a square root.

The equations of motion are:
\bea
\label{eom1}
\frac{dt}{ ds} &=& -2\lambda (p-V); \\
\label{eom2}
\frac{dp}{ ds} &=& 0; \\
\label{eom3}
\frac{dR}{ ds} &=& 2\lambda P; \\
\label{eom4}
\frac{dP}{ ds} &=&\lambda \frac{m^2(p-V)}{R^2}.
\eea
These expressions are equivalent to the dynamics described by (\ref{basic}), which can be seen as follows. The second equation tells us that $p = -E$ is a constant of motion. Using the first equation, we find from (\ref{eom3}) and (\ref{eom4}) respectively that
\bea
\label{eoma}
\dot{R} &=& \frac{dR}{ds}\frac{ds}{dt} = \frac{P}{E+V},\\
\label{eomb}
\dot{P} &=& \frac{dP}{ds}\frac{ds}{dt} = -\frac{m^2}{2R^2}.
\eea
Taking the $t$-derivative of (\ref{eoma}) and combining this with (\ref{eomb}), one obtains an expression which is equivalent to the $t$-derivative of (\ref{basic}).

This establishes the classical, canonical theory that  we shall quantize in the next section.

\section{Quantum theory}
In this section we establish a Hilbert space of functions $\Psi(t,R)$, and represent the phase space variables $(t, p, R, P)$ as operators on this space. We then write (\ref{sh}) as a quantum constraint $\hat{h}$ in terms of these operators and look for solutions.

\subsection{Auxiliary Hilbert space}
The polymer Hilbert space is built upon the real line with discrete topology (i.e. the Bohr compactification of the real numbers $x \in \mathbb{R}_{\mbox{\scriptsize Bohr}}$) \cite{poly, HL}.  Position eigenstates are defined as:
\be
\psi_{x}(R) = \left\{
\begin{array}{ll}
1, & R = x \\
0, & R \ne x
\end{array} \right. .
\ee
The basis is uncountable and therefore unseparable, but we will find that solutions to the constraint $\hat{h}$ are defined on a superselection sector which has a countable basis and is separable.

The polymer inner product is:
\be
\left(\psi_{x}, \psi_{x'}\right) = \delta_{x, x'},
\ee
where $\delta_{x, x'}$ is the Kronecker delta.

A general, time-dependent state is written as:
\be
\Psi(t,R) = \sum_x c_x(t) \psi_x(R),
\ee
where each $c_x(t)$ is a time-dependent coefficient for the basis state $\psi_x(R)$. Using the inner product we find that the norm squared is:
\be
(\bar{\Psi}(t,R), \Psi(t,R)) = \sum_x \bar{c}_x(t) c_x(t) ,
\ee
which notably depends upon time.

We choose the usual representation for the first pair of operators where $\hat{t}$ acts by multiplication and $\hat{p}=i \partial_t$ acts as a partial derivative
\be
\hat{t}\Psi = t\Psi, \qquad \hat{p} \Psi = i \sum_x \dot{c}_x \psi_x
\ee
The second pair of operators requires more care. Since we have position eigenstates, the action of the position operator is simply
\be
\hat{R} \psi_x = x \psi_x.
\ee
However, the usual definition of the momentum operator $\hat{P}$ as a continuum partial derivative is not well defined because space is discrete. In order to construct a well defined $\hat{P}$, we introduce a finite translation operator $\hat{U}_{{\mu}}:=e^{i\hat{P}{\mu}}$ which shifts a position eigenstate by $\mu \in \mathbb{R}$:
\be
\hat{U}_{{\mu}} \psi_{x}(R) = \psi_{x}(R+{\mu}) = \psi_{x-{\mu}}(R),
\ee
A well-defined momentum operator which depends upon the spacing ${\mu}$ is given by:
\be
\hat{P} = \frac{1}{i {\mu}} \left(\hat{U}_{\frac{\mu}2} - \hat{U}_{\frac{\mu}2}^\dagger \right).
\ee
Squaring this yields:
\be
\hat{P}^2 = \frac{1}{\mu^2}  \left(2 - \hat{U}_{\mu} - \hat{U}_{\mu}^\dagger \right).
\ee

In principle, $\mu$ is free to vary in both time and space. However in order to keep the equations manageable, we choose $\mu$ to be an arbitrary fixed constant. This amounts to partially choosing a lattice.  Since $\mu$ will enter the constraint equation only in the $\hat{P}^2$ term, a choice of $\mu$ picks out the superselection sector; only the coefficients $c_x$ and eigenstates $\psi_x$ which correspond to points $x$ on a regular $\mu$-spaced lattice will have an effect on each other. In general, a lattice is the following set of points
\be
L_{\sigma} = \{ x = \sigma + k \mu \ | \ k \in \mathbb{Z} \}, \\
\ee
where the choice of $0 \le \sigma < \mu$ picks the superselection sector. Here we study the theory on two of these lattices corresponding to the choices $\sigma = 0, \frac{1}{2}$; one which contains the point $x=0$ and one which does not.

Notice that the lattices defined above do not have a bound on $x$. This is necessary for the $\hat{P}^2$ operator to be well-defined (with the same definition) on all position eigenstates \footnote{Alternatively one could consider only the points $x \ge 0$ and alter the definition of $\hat{P}^2$ at the first lattice point as done in \cite{KL}}. This means that in principle, we are allowing for negative eigenvalues of the position operator, but since the $\hat{R}$ operator comes from a positive semi-definite radial variable $R \ge 0$, we will study solutions with support on the $x \ge \sigma$ lattice points only. There are conserved currents coming from the constraint equation, and we apply boundary conditions which conserve the corresponding charges on the $x \ge \sigma$ portion of the lattice.

We now define the auxiliary Hilbert space (without boundary conditions):
\be
\mathcal{H}_{\mbox{\tiny aux}}^\sigma := \left\{ \left. \Psi(t,R) = \sum_x c_x(t) \psi_x(R) \ \right| \ \sum_{x \in L_\sigma} c^*_x(t) c_x(t) \le \infty \right\}.
\ee
This is a general space of functions within which the operators and inner product are well-defined. The solutions to $\hat{h}$ will comprise a subset of this space, and appropriate boundary conditions will be introduced to define the charge-conserving portion of this solution space.

\subsection{Hilbert space of constraint solutions}
In this subsection we define the physical Hilbert space of solutions to $\hat{h}\Psi = 0$, or equivalently
\be
\label{opeq}
-\left[\hat{p}-V(\hat{R})\right]^2\Psi(t,R) + \left[\hat{P}^2 + m^2\right]\Psi(t,R) = 0.
\ee
Using the definitions in the previous subsection, the constraint equation for states in $\mathcal{H}_{\mbox{\tiny aux}}$ can be written as:
\be
\sum_x \left[ \left(\ddot{c}_x - \frac{im^2}{x} \dot{c}_x + \left(\frac{2}{\mu^2}+m^2- \frac{m^4}{4x^2}\right)c_x \right) \psi_x -\frac{c_x}{\mu^2}\left(\psi_{x+\mu}+\psi_{x-\mu} \right)\right] = 0.
\ee
If we shift the summation for the last two terms, and notice that the above condition must hold independently at each point $x$, we arrive at an equation of motion for the coefficients:
\be
\label{eom}
\ddot{c}_x - \frac{im^2}{x} \dot{c}_x + \left(\frac{2}{\mu^2}+m^2- \frac{m^4}{4x^2}\right) c_x - \frac{1}{\mu^2} \left(c_{x+\mu} + c_{x-\mu} \right) = 0.
\ee
Notice that some of the terms in the above equation are divergent at $x=0$ on the lattice with $\sigma = 0$, but only in this case.

In order to fix boundary conditions, we first note that there are two conserved charges. These are the polymer analogs to the Klein-Gordon (KG) inner product and the energy form defined in \cite{HKK}. Given two states $\Phi = \sum_x b_x \psi_x$ and $\Psi = \sum_x c_x \psi_x$ we define the following bilinear forms:
\bea
\label{KG}
q(\Phi, \Psi) &=& \frac{1}{2} \sum_{x \ge \sigma} \left( i \left( \bar{b}_x \dot{c}_x - \dot{\bar{b}}_x c_x \right) + \frac{m^2}{x} \bar{b}_x c_x \right), \\
\label{energy}
e(\Phi, \Psi) &=& \frac{1}{2} \sum_{x \ge \sigma} \left( \left( \dot{\bar{b}}_x \dot{c}_x +\frac{1}{\mu^2} \left( {\bar{b}}_{x+\mu} - {\bar{b}}_x \right) \left(c_{x+\mu} - c_x \right) +\left(m^2 - \frac{m^4}{4x^2} \right)\bar{b}_x c_x \right) \right),
\eea
where $q$ is the KG inner product and $e$ is the energy.

One can check that $e$ and $q$ are conserved on the $x \ge \sigma$ portion of the lattice under the dynamics defined by (\ref{eom}) provided that one of the following boundary conditions holds:
\bea
\label{c1}
\mbox{condition 1}: \quad && b_\sigma = c_\sigma = 0; \\
\label{c2}
\mbox{condition 2}: \quad && b_\sigma = b_{\sigma - \mu}, \quad c_\sigma = c_{\sigma - \mu}.
\eea
Having a choice of boundary conditions which conserve the charges is different than in the continuum theory where only a single option is available. This is a direct result of discretization: the continuum boundary condition defines the limiting form of the wavefunction at small $R$, while the polymer boundary conditions involve the value of $\Psi(t,R)$ at one or two distinct points using (\ref{c1}) or (\ref{c2}) respectively. We find numerically (as described below) that in the limit of $\mu \to 0$, the two polymer boundary conditions agree $c_{\sigma - \mu} = c_\sigma = 0$, so that there is no contradiction with the continuum theory.

We now define the Hilbert space of solutions to the quantum constraint by applying additional conditions to the more general $\mathcal{H}_{\mbox{\tiny aux}}^\sigma$:
\be
\mathcal{H}_{I}^\sigma = \left\{ \left. \Psi \in \mathcal{H}_{\mbox{\tiny aux}}^\sigma \ \right| \ \hat{h}\Psi = 0, \ \beta_I = 0 \right\},
\ee
where the subscript $I = 1, 2$ labels the choice of boundary conditions:
\be
\beta_1:=c_\sigma, \qquad \qquad \beta_2:= c_\sigma - c_{\sigma - \mu}.
\ee
States in this Hilbert space satisfy the quantum constraint and preserve the $q$ and $e$ charges on the $x \ge \sigma$ portion of the lattice.
With this definition we can move forward and solve for these states explicitly.

\section{Solutions}
In this section we find solutions to the constraint $\hat{h} \Psi = 0$. We first look at time-independent solutions and then study the dynamics.

\subsection{Stationary states}
Let us now find explicitly the states $\Psi_E \in \mathcal{H}_{I}^\sigma$ which are eigenstates of the Hamiltonian $\hat{H} \Psi = E \Psi$ (where $\hat{H} = i \partial_t$ in this representation), within the solution space of the quantum constraint $\hat{h} \Psi = 0$. These are found using the following ansatz for the time-dependent part of the states:
\be
\label{ansatz}
c_x(t) = C_x e^{-iEt},
\ee
where each $C_x \in \mathbb{C}$ is independent of $t$. Putting this ansatz in (\ref{eom}) results in the following difference equation for the constants \footnote{Similar difference equations occur in polymer quantum mechanics, e.g. (V.3) in \cite{poly} and (43) in \cite{gks}.}:
\be
(2 - \mu^2 \Delta - f_k) C_k - C_{k-1} - C_{k+1} = 0 , \label{eq:recurs1}
\ee
where we have used the integer label $k = (x-\sigma)/\mu$ for points on the lattice and defined
\be
\Delta \equiv E^2 - m^2, \qquad \qquad f_k \equiv \mu^2 m^2 \left(\frac{E}{x} + \frac{m^2}{4x^2}\right) .
\ee

On a positive energy eigenstate of the form (\ref{ansatz}), the KG inner product (\ref{KG}) and energy form (\ref{energy}) are
\bea
q(\Psi, \Psi) &=& \frac{1}{2}\sum_{x \ge \sigma} \left(E+\frac{m^2}{2x} \right) |C_x|^2 , \\
e(\Psi, \Psi) &=& \frac{1}{2}\sum_{x \ge \sigma} \left[ \left(E^2 + m^2 - \frac{m^4}{4x^2} \right) |C_x|^2 + \frac{1}{\mu^2} \left|C_{x+\mu} - C_x\right|^2 \right].
\eea
Notice that each summand of $q$ is positive semi-definite for $E>0$. %and so can be used as a probability summand on the space of positive energy eigenstates.
The energy summands may become negative at small values of $x$, but are positive definite for $x \ge \frac{m}{2}$. In the continuum theory, the fall-off conditions at small $R$ guarantee that the energy form is positive definite. Note however that only the total sums $q$ and $e$ are physical observables while the summands themselves are not.

Using (\ref{eq:recurs1}) and either boundary condition, the energy form is found to be
\be
e(\Psi, \Psi) = E q(\Psi, \Psi),
\ee
so that if the states are normalized to make $q=1$, then the bilinear form $e$ measures energy.

The difference equation cannot be solved analytically, but using the fact that $\lim_{k \to \infty}f_k = 0$, we can solve the equation at large $k$:
\be
\label{largek}
C_k = A^{\pm k}, \qquad A \equiv 1 - \frac{\mu^2 \Delta}{2} + \sqrt{\left(1 - \frac{\mu^2 \Delta}{2} \right)^2 - 1}.
\ee
These large $k$ solutions have different behaviours for different values of energy at fixed $m$ and $\mu$, and this characterizes the states.

\begin{description}
\item [Bound states] For $|E| < m$ we have that $|A| > 1$. Normalizable states require the negative choice of sign $C_k = A^{-k}$, giving states which go to zero at large $k$.
\item [Scattering states] For energies in the range $m \le |E| < \sqrt{\frac{4}{\mu^2}+m^2}$ we get that $|A| = 1$ so that these solutions are neither normalizable nor divergent. Energy may be of either sign, and the bounds go to $\pm \infty$ in the limit of $\mu \to 0$. %The negative energy states are antiparticles.
\item [Non-existent states] For the range $|E|  \ge \sqrt{\frac{4}{\mu^2}+m^2}$ we have that $|A| < 1$ which requires a positive choice of sign $C_k = A^k$ for normalizable solutions. These states have no counterpart in the continuum theory since in the $\mu \to 0$ limit, bound and scattering states cover the entire energy spectrum. In the polymer theory, we do not find any states at these energies which satisfy either boundary condition.
\end{description}

Since the difference equation cannot be solved analytically we resort to numerical techniques. A time independent solution is defined by knowing all of the non-zero coefficients $C_k$, and these can be found using a shooting method. To do so, one checks many different values of $E$ to search for those which satisfy the boundary conditions. Given a particular `guess' $E$, one begins with the solution (\ref{largek}) at some $k$ which satisfies $k \gg \mu E$ and $k \gg m/2$, then iterates downward according to the full difference equation (\ref{eq:recurs1}). After the coefficients are found all the way down to the first lattice point, one checks whether one of the boundary conditions (\ref{c1}, \ref{c2}) is satisfied. If the `guess' at $E$ results in a solution which satisfies the choice of boundary condition, this state is recorded as an eigenstate. Otherwise it is discarded as lying outside the space of solutions. (All of the results in this section were confirmed using an alternative numerical method based on continued fractions \footnote{A brief description of the continued fraction method is the following. One may write the difference equation (\ref{eq:recurs1}) as
\be
F_k C_k-C_{k-1} = C_{k+1},
\ee
where $F_k \equiv 2-\mu^2 \Delta - f_k$. Then defining $\alpha_k = C_{k+1}/C_k$ we have the recursion relation
\be
\alpha_{k-1} = (F_k - \alpha_k)^{-1}.
\ee
Now, at large $k$ where the asymptotic solution (\ref{largek}) holds, we have $\alpha_k \approx A$. Iterating downward from here, one can check whether: $F_1 = \alpha_1$ for the boundary condition $C_0 = 0$; $F_1-1 = \alpha_1$ for the boundary condition $C_0 = C_{-1}$.
}.)

\subsubsection{Bound states}

Let us first review the results of the continuum theory for comparison where bound states were found only for $m<1$; there are no continuum solutions for $m\ge 1$ which satisfy the boundary conditions. The spectrum for $m<1$ is
\be
\label{conspec}
E_n = m \frac{2(\lambda + n)}{\sqrt{m^4+4(\lambda + n)^2}}, \qquad \qquad \lambda = \frac{1}{2} + \frac{1}{2} \sqrt{1-m^4},
\ee
where $n$ is an integer greater than or equal to zero. Notably, the entire spectrum is positive.

In the polymer theory we also find bound states, but without any restriction on $m$. These exist on both choices of lattice ($\sigma = 0$ and $\sigma = \mu/2$) and for both choices of boundary conditions. For $m<1$ the continuum spectrum is recovered in the limit of $\mu \to 0$.

\begin{figure}
\centering
\subfigure[Probability density ($q =: \sum_x q_x$).]
{\includegraphics[scale=0.5]{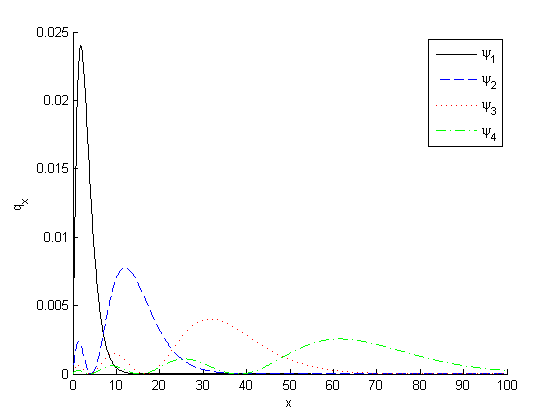}}\qquad
\subfigure[Energy density ($e =: \sum_x e_x$).]
{\includegraphics[scale=0.5]{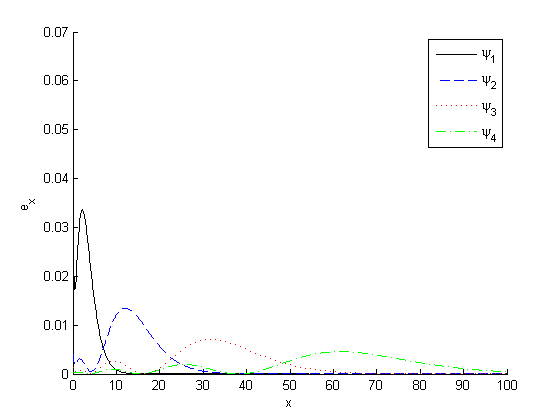}}
\subfigure[First ten lattice points of $q_x$.]
{\includegraphics[scale=0.5]{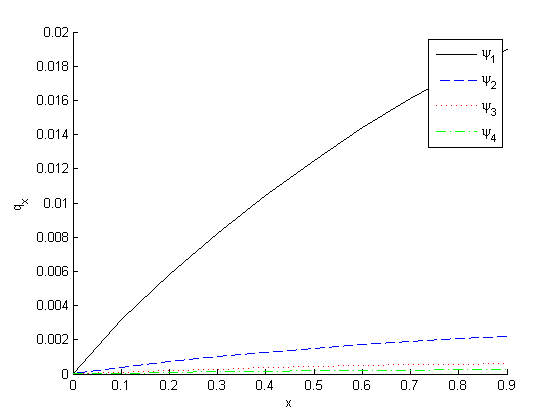}}\qquad
\subfigure[First ten lattice points of $e_x$.]
{\includegraphics[scale=0.5]{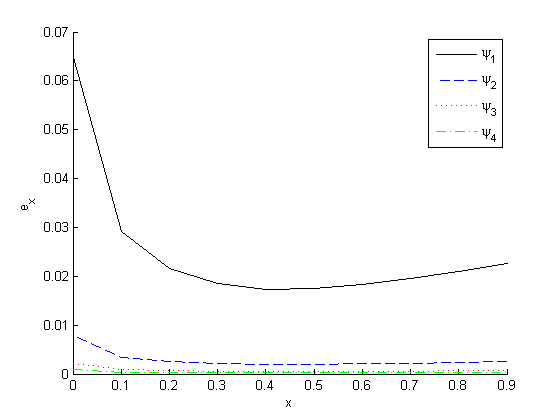}}
\caption{\label{ef} Probability and energy densities of the lowest four bound eigenfunctions for $m = 0.9$ satisfying $C_0=0$ on the lattice with $\sigma = 0$ and $\mu = 0.1$. The corresponding energies are $(0.8067, 0.8785, 0.8909, 0.8950)$. Notice in (c) and (d) that the energy density grows near the origin while the probability density goes to zero.}
\label{fig:bc1}
\end{figure}

\begin{figure}
\centering
\subfigure[Probability density ($q =: \sum_x q_x$).]
{\includegraphics[scale=0.5]{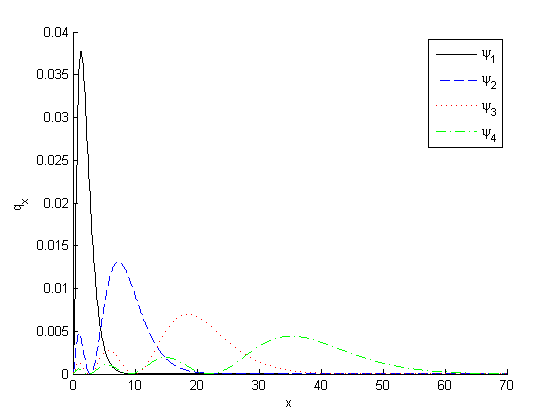}}\qquad
\subfigure[Energy density ($e =: \sum_x e_x$).]
{\includegraphics[scale=0.5]{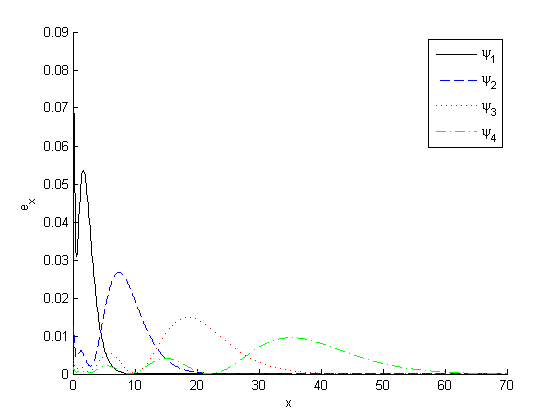}}
\subfigure[First ten lattice points of $q_x$.]
{\includegraphics[scale=0.5]{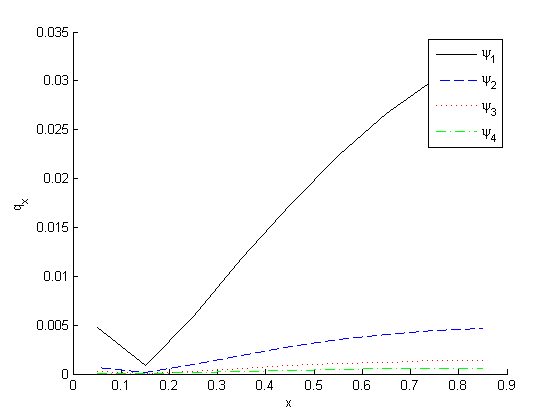}}\qquad
\subfigure[First ten lattice points of $e_x$.]
{\includegraphics[scale=0.5]{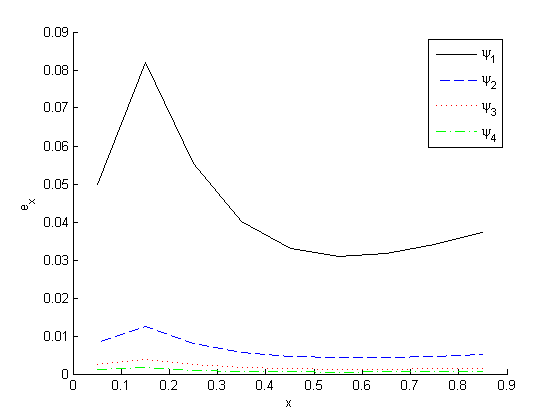}}
\caption{\label{ef2} Probability and energy densities of the lowest four bound eigenfunctions for $m = 1.1$ satisfying $C_{-1} = C_0$ on the lattice with $\sigma = \frac{1}{2}$ and $\mu = 0.1$. The corresponding energies are $(0.8995, 1.0468, 1.0764, 1.0868)$. Notice in (c) and (d) that there are cusps at $x = \sigma + \mu$ in the probability and energy densities due to the choice of boundary condition. The coefficients themselves do not have a cusp.}
\end{figure}

In Fig. \ref{ef} the densities associated to the KG and energy forms are plotted for the lowest four bound eigenfunctions for $m = 0.9$ satisfying $C_0=0$ on the lattice with $\sigma = 0$ and $\mu = 0.1$. Since the energies are all positive, the KG inner product is positive semi-definite and may be interpreted as the probability density. The two densities have a similar profiles outside of the first ten or so lattice points; as the eigenvalue increases, both densities move further from the origin and become more spread. Increasing $m$ tends to focus the densities closer to $R=0$, while the solutions are insensitive to the choice of $\mu$. A large $\mu$ gives a coarse approximation to the same curves. For comparison, we also provide plots in Fig. \ref{ef2} of the lowest four bound eigenfunctions for $m=1.1$ satisfying $C_{-1} = C_0$ on the $x = \mu k + \frac{1}{2}$ lattice.

Using the probability density, we can calculate radial expectation values and check where stationary states lie in relation to their Schwarzschild radii. We note here that such values are not physical observables since $\hat{R}$ does not commute with $\hat{h}$, and will comment more on this in the discussion. In Table \ref{rad} the radial expectation values corresponding to the plots in Fig. \ref{ef} are compared with the Schwarzschild radii. None of these expectation values are within the corresponding Schwarzschild radius, getting further out as the eigenvalue increases. The radial expectation values lie within the Schwarzschild radius only for $m\ll 1$ \cite{HKK}, but we are unable to probe low enough values due to numerical limitations.

%\begin{figure}
%\includegraphics[scale = 0.67]{bound.png}
%\label{bound}
%\caption{Coefficients defining the first four bound eigenfunctions $m = 0.9$ satisfying $C_0=0$ on the lattice with $\sigma = 0$ and $\mu = 0.1$.}
%\end{figure}

\begin{table}[]
\begin{tabular}{r || c | c | c | c}
$\langle R \rangle_n$ & 3.0839 & 13.9558 & 33.0511 & 60.3732 \\
\hline
$2E_n$ & 1.6134 & 1.7570 & 1.7817 & 1.7900
\end{tabular}
\caption{\label{rad} Radial expectation values $\langle R \rangle_n$ of lowest four energy eigenstates compared with the Schwarzschild radii $2E_n$. Here we used $m = 0.9$ and the boundary condition $C_0=0$ on the lattice with $\sigma = 0$ and $\mu = 0.1$.}
\end{table}

\begin{figure}
\centering
\subfigure[Boundary condition $c_0 = 0$ on the lattice with $\sigma = 0$.]
{\includegraphics[scale=0.5]{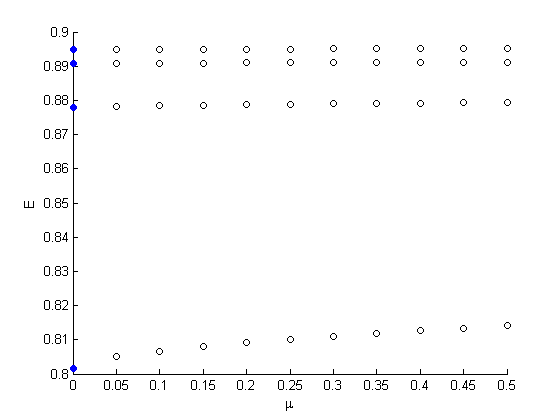}}\\
\subfigure[Boundary condition $c_0 = 0$ on the lattice with $\sigma = \mu/2$.]
{\includegraphics[scale=0.5]{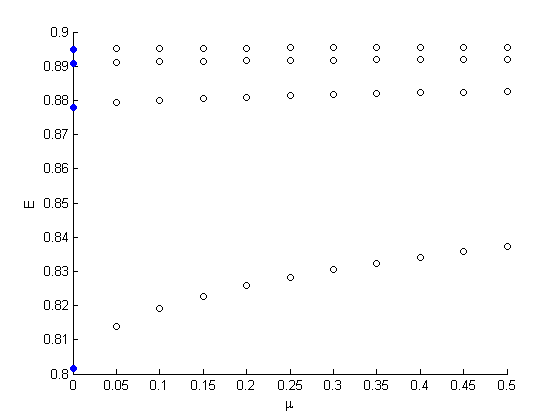}}\quad
\subfigure[Boundary condition $c_0 = c_{-1}$ on the lattice with $\sigma = \mu/2$.]
{\includegraphics[scale=0.5]{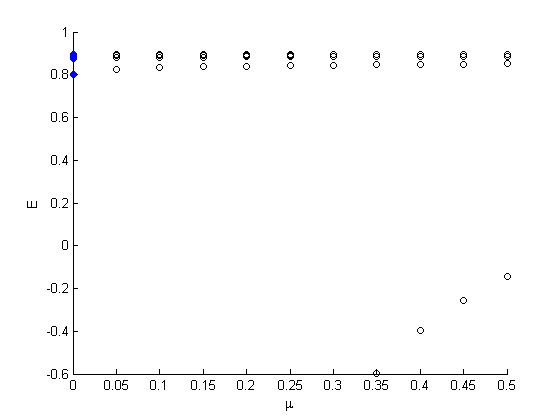}}
\caption{Polymer energy spectra for rest mass $m=0.9$ with different lattice spacings $\mu$ ranging from $0.05$ to $0.5$. The spectrum resulting from Dirac quantization is shown in blue filled circles at $\mu = 0$ (the y-axis).}
\label{m=0.9}
\end{figure}

Figure \ref{m=0.9} shows the spectrum for $m=0.9$ at various spacings $\mu$. It is interesting that for the boundary condition (\ref{c2}) the ground state may have a negative energy eigenvalue for a certain range of $\mu$. As $\mu$ goes to zero, this negative eigenvalue goes below $-m$ and leaves the spectrum, and the next lowest eigenstate becomes the ground state. When $m<1$, for any choice of lattice or boundary condition the polymer spectrum agrees with the continuum spectrum in the limit of $\mu \to 0$. Note that one cannot use boundary condition (\ref{c1}) on the $\sigma=0$ lattice due to divergences in the difference equation which prevent finding $c_{-1}$.

\begin{figure}
\centering
\subfigure[Boundary condition $c_0 = 0$ on the lattice with $\sigma = 0$.]
{\includegraphics[scale=0.5]{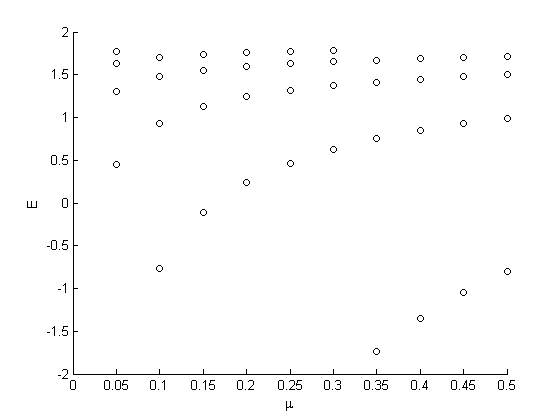}}\\
\subfigure[Boundary condition $c_0 = 0$ on the lattice with $\sigma = \mu/2$.]
{\includegraphics[scale=0.5]{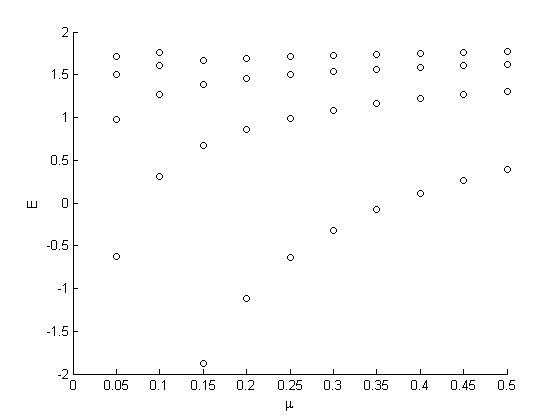}}\quad
\subfigure[Boundary condition $c_0 = c_{-1}$ on the lattice with $\sigma = \mu/2$.]
{\includegraphics[scale=0.5]{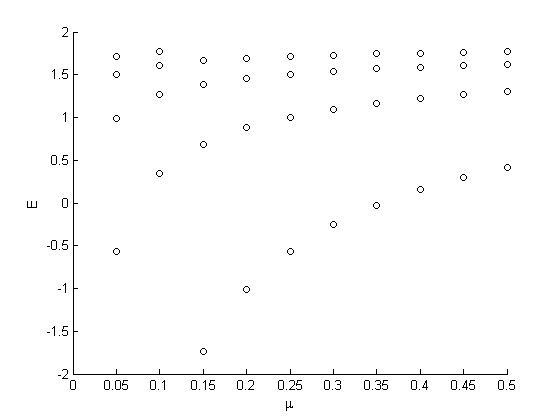}}
\caption{Polymer energy spectra for rest mass $m=2$ with different lattice spacings $\mu$ ranging from $0.05$ to $0.5$. No such eigenvalues exist in the continuum theory.}
\label{m=2}
\end{figure}

The polymer boundary conditions have no dependence on $m$, and so has bound state eigenvalues for $m>1$ which have no counterpart in the continuum theory. As the spacing $\mu \to 0$, the eigenvalues in the $m>1$ spectrum continue to fall for smaller and smaller $\mu$, and do not converge to any fixed values, so that there is no contradiction between the polymer and continuum theories. See Fig. \ref{m=2} which shows the spectrum for $m=2$ at various spacings $\mu$. For $m>1$ the ground state may have negative energy with either boundary condition, at certain values of $\mu$, but the first excited state and above are found to be positive in all cases tested numerically.

\subsubsection{Scattering states}
In the continuum theory, scattering solutions are found only when $m<1$ as is the case for bound states. The solutions are given by Coulomb wave functions and have a continuous energy spectrum for both positive and negative energies $m < |E| < \infty$.

From the limiting form of the solution at large $k$ (\ref{largek}), scattering solutions in the polymer theory are
\be
C_k \sim e^{\pm i \omega k}
\ee
where $\cos \omega = 1 - \frac{\mu^2 \Delta}{2}$. At large $k$ these solutions are ingoing and outgoing modes.

\begin{figure}
\centering
\subfigure[Solution which is $e^{i\omega k}$ at large $k$.]
{\includegraphics[scale=0.5]{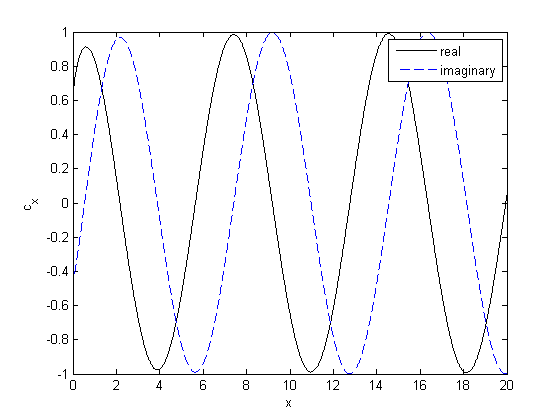}}\qquad
\subfigure[Solution which is $e^{-i\omega k}$ at large $k$.]
{\includegraphics[scale=0.5]{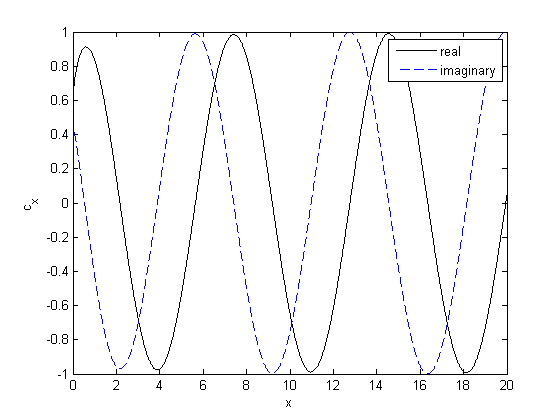}}
\caption{Real and imaginary parts of the coefficients describing typical scattering solutions. Here $m=\frac{1}{2}$, $E = 1$, and the lattice is $x = k \mu$. Notice the real parts of $C_0$ are the same for both solutions, while the imaginary parts have the same magnitude but opposite sign.}
\label{scatter}
\end{figure}

On the lattice with $\sigma = 0$ and using boundary condition (\ref{c1}), one finds that the coefficient $C_0^+$ obtained from the shooting method starting with $e^{i \omega k}$ at large $k$, is the complex conjugate of the coefficient $C_0^-$ obtained from the solution which is $e^{- i \omega k}$ at large $k$, e.g. $C_0^+ = \overline{C}_0^-$ for any choice of $E$ in the scattering range. See fig. \ref{scatter} for an example. Therefore the following linear combination satisfies (\ref{c1}) for any scattering energy:
\be
\Phi_E = \Psi^+ - e^{2i \arg C_0^+} \Psi^-,
\ee
where $\Psi^\pm$ are the solutions with $C_k \sim e^{\pm i \omega k}$ at large $k$. A similar argument applies for other choices of lattice and boundary condition (\ref{c1}, \ref{c2}). This argument shows that the scattering spectrum is continuous for either choice of boundary condition and any choice of $\sigma$, as one would expect. Furthermore, at large $k$ the scattering solutions are a linear combination of ingoing and outgoing modes indicating a bounce at the origin with a particular phase shift as in the continuum theory \cite{HKK}.

\subsection{Time dependent states}
In the previous subsection we obtained the complete energy spectrum, and each value in the spectrum labels a solution to $\hat{h} \Psi_E = 0$. Consider a Hilbert space defined using the KG inner product on such states. If we restrict to the positive energy states, the KG density is positive semi-definite and may be used as a measure of probability. In this Hilbert space, an arbitrary state is written as
\be
\Phi = \sum_E \beta_E(t) \Psi_E.
\ee
%The question remains whether this space is complete.

In principle, one could build an electronic database which represents the Hilbert space described in the previous paragraph. This would involve storing the coefficients of a large number of energy eigenstates for a fixed $m$, on a given lattice and for a given choice of boundary condition. One could then attempt to construct semiclassical initial data and study the evolution of such data by calculating the equations of motion for the coefficients $\beta_E(t)$. However, we have not taken up this task here. Rather, we pursue a numerical solution to (\ref{eom}) based on the finite differences naturally defined by the polymer theory. From this angle, we can ask whether the dynamics we find can be described using positive energy eigenstates only.

The equation (\ref{eom}) is well-suited to numerics since the terms coming from $\hat{P}^2$ effectively define a second order finite difference derivative with local error terms on the order of $\mathcal{O}(\mu)$. The lattice is also fixed by the polymer theory, up to choices of $\sigma$ and $\mu$, and the boundary conditions necessary to conserve the $q$ and $e$ charges have been given in (\ref{c1}, \ref{c2}). In fact, the only thing left to do is to choose a method of time integration. For this purpose we use Heun's method, which is second order accurate in the time step $dt$. Stability (the CFL condition \cite{CFL}) requires that we limit the number of lattice spaces that data can move in a single time step, which is done by setting $dt = \frac{\mu}{2}$.

In the code, we specify initial coefficients $c_x$ and initial velocities $\dot{c}_x$, then integrate forward in time according (\ref{eom}). In a collapse problem such as this, one often encounters difficulty near the origin where the data may become large relative to values elsewhere. Summing large numbers near $R=0$ can yield significant round-off errors and lead to numerical instability. Because of this it is often useful to impose a fall-off condition on the solution in order to improve the stability of the code.

In order to improve stability we borrow from the continuum theory, since the numerical solutions converge to continuum solutions in the $\mu \to 0$ limit. In \cite{HKK} it was found that the fall-off behaviour at small $R$ must be $R^{\lambda_+}$ where $\lambda_\pm = \frac{1}{2} \pm \frac{1}{2} \sqrt{1 - m^4}$. There exists another solution in the continuum theory which falls-off as $R^{\lambda_-}$, but this one does not satisfy the boundary condition. For $m<1$ the `good' solution goes to zero while the `bad' solution diverges, but for $m \ge 1$ both solutions have the same fall-off up to oscillatory terms. For $m<1$, we improve \footnote{The authors are grateful to Andrei Frolov for this suggestion.} the stability of our code by writing the coefficients as $b_x:= R^{\lambda_+} c_x$ and solving for the $b_x$ data at each time step. We monitor stability by tracking the charges $q$ and $e$ throughout the dynamics. Decreasing the lattice spacing and time step improves the conservation while increasing run-time.

We study semiclassical ingoing initial data of the following form:
\be
c_x(0) \sim \sqrt{2x}e^{-\frac{(x-x_0)^2}{2w^2}}, \qquad \qquad \dot{c}_x(0) \sim \frac{c_{x+\mu}-c_{x}}{\mu} ,
\ee
with the proportionality constants chosen to fix $q=1$. For such data, the probability density is an inward-moving (up to some small outgoing corrections) Gaussian pulse of width $w$ centred at $x_0$, and the KG density is positive semi-definite. We choose the proportionality constants to normalize $q = 1$ so that $e$ is a measure of the energy.
%We looked at other forms of initial data\footnote{We also looked at initial coefficients such that each $c_x$ is a point on a Gaussian curve, as well as vanishing initial velocity which results in both outward and inward moving components.}, but the results are similar for any choice. For all choices of parameters which give a stable numerical evolution, we find that the initial data is well-within $2e$, which represents a mean value of the associated Schwarzschild radius (recalling that eigenstates of different energies are summed over to represent such a state).

In the following cases, we study the dynamical behaviour for different values of the rest mass $m$ on the lattice with $\sigma = 0$. Animations of the KG and energy densities for each of the solutions discussed below can be found at \href{http://ion.uwinnipeg.ca/~gkunstat/Polymer/}{http://ion.uwinnipeg.ca/~gkunstat/Polymer/}.
\begin{description}
\item[$m \lsim 0.1$] For small values of rest mass we find that the wavefunction collapses inward, bounces at the origin, then moves outward toward infinity. Animations of the KG and energy densities for $m = 0.01$ with $\mu = 10^{-3}$ can be found at the web address given above. The energy density shows two peaks which bounce off the origin while maintaining their shape. The KG density begins as a single pulse which splits during the bounce leaving a completely negative region which coincides with the outer energy peak. Due to this negative region, we are hesitant to refer to the KG density as a probability density. The total energy is $5.05\cdot10^4$ in Planck units with deviations of less than $0.3\%$, while $q$ remains fixed at $1$ with deviations of less than $6 \cdot 10^{-5}\%$.

\item[$0.1 \lsim m < 1$] As the rest mass approaches the Planck mass from below, we find we find that the wavefunction continues to bounce off the origin, but now exhibits self-interaction (dispersion) throughout the process. The increasing spread of the densities implies the state is moving away from semiclassicality and becoming more quantum as $m$ increases. As an example, we have generated animations of the KG and energy densities for $m=0.5$. Here the total energy is $25.2$ with deviations of less than $1.6\%$, while $q$ remains normalized within $8 \cdot 10^{-5}\%$.

\item[$m \ge 1$] The numerics become increasingly unstable as $m$ exceeds the Planck mass, and may crash for certain choices of the other parameters. Given that this polymer theory is effectively a finite difference approximation of the continuum, we expect the numerical solution to follow the continuum behaviour more closely as $\mu \to 0$. Since the continuum theory does not have solutions for this range of $m$, we expect the numerics to fail for small $\mu$. These numerical difficulties are lessened for large values of $\mu$ which moves the first lattice point further from the origin where the potential diverges \footnote{Choosing a lattice with $\sigma > 0$ has a similar effect.}. For example, with $m=1$ and $\mu = 10^{-4}$, the code becomes unstable when the pulse reaches the origin, while it exhibits a bounce (with some noise near the origin) when $\mu = 0.1$. See for example the animations of KG and energy densities for $m=1$ with $\mu = 0.1$. Here the total energy is $10.0$ but undergoes a large deviation of $\sim 9 \%$ during the bounce, while the KG inner product remains conserved within $0.6\%$.
\end{description}

For all parameter ranges for which the code is stable, we find generically that the probability density becomes negative locally at some time during the evolution, and these negative portions do not go away as evolution continues. It is possible that such an effect could be due to numerical inaccuracy. However, negative KG density for ingoing Gaussian initial data is seen even for very small values of $m$ where the numerics are optimal. The implication is that the dynamics we find numerically cannot be described in terms of positive energy eigenstates, i.e. these states $\Psi(t,R)$ are not within the Hilbert space discussed at the beginning of the section. Whether negative eigenstates are a requisite component of any bouncing state is an open question.

\section{Discussion}
We studied the polymer quantum mechanics of a self-gravitating thin shell. The energy spectrum was found by solving for stationary eigenstates using the shooting method. When $m <1$, the spectrum matches the continuum spectrum found in \cite{HKK} as the polymer scale $\mu \to 0$. We find two independent choice of boundary conditions which conserve energy and the KG inner product, and in the limit $\mu \to 0$ these are both satisfied and are consistent with the ($m<1$) continuum boundary condition. For finite values of $\mu$ we find that there may be negative energies associated with bound states which is something not seen in the continuum theory. For $m \ge 1$ where no continuum solutions exist, we find that the bound state spectrum does not converge to fixed values as $\mu \to 0$, so that there is no contradiction.

In the classical theory, all initial conditions lead to black hole formation. Bound states are not present, so it is very interesting to find them in the quantum theory. Replacing the classical particle (shell) by a quantum wavefunction has allowed for a superposition of ingoing and outgoing solutions with total energy $|E|<m$ which satisfy the boundary condition. For positive energies, these states have positive definite KG density. If we interpret this as the probability density, these are highly quantum states with a large uncertainty in position. Scattering states for $|E| \ge m$ are explicitly a combination of ingoing and outgoing states at large $x$. Since charges are conserved by these states, no energy is lost down a singularity at $R=0$, indicating a unitary quantum evolution.

A question this work brings to light is the meaning of states in quantum gravity. Each state is a functions of both time and space, representing a superposition of shell spacetimes. In this formalism, how can we extract semiclassical information and recover a single notion of spacetime? The physical observables seem limited to the total energy, which defines the location of a Schwarzschild radius but does not indicate the shell location. If one wants to learn more about the spacetime, it seems something more is needed. To this end we considered the radial expectation value to represent the position of the shell. Even classically, the value of the radius at a particular time is a slicing (gauge) dependent quantity. In quantum gravity we may need to be open to such gauge-dependent observables. %It would be interesting to study the mathematical consistency of gauge-covariant (rather than invariant) observables.

We studied the dynamics of an ingoing Gaussian probability density, taken to represent a semiclassical ingoing shell for small $m \ll 1$, and becoming increasingly quantum (with a larger spread in probability density) as $m$ approaches the Planck mass from below. The probability density is positive semi-definite only for positive energy eigenstates. For all $m$ less than one Planck mass, we find that the wavepacket bounces off the origin, but picks up regions of negative probability density. This implies that the bouncing states we find cannot be represented as a sum over positive energy eigenstates. This suggests that negative energy eigenstates may play an important role in quantum singularity resolution. It is not clear whether or not the standard physical interpretation of these states as positive energy solutions moving background in time is relevant in the present context. In any case, the present model seems to be an ideal theoretical laboratory for studying this and other important issues in quantum gravity.

\acknowledgments{GK thanks Andrei Frolov for helpful conversations and Simon Fraser University for its kind hospitality during part of the completion of this work. The authors are also grateful to Viqar Husain and Jorma Louko for helpful discussions regarding observables in quantum gravity. This work was funded in part by the Natural Sciences and Engineering Research Council of Canada as well as the Atlantic Association for Research in the Mathematical Sciences.  Support was also provided by the Perimeter Institute for Theoretical Physics (funded by Industry Canada and the Province of Ontario Ministry of Research and Innovation).}


\begin{thebibliography}{5}%
\makeatletter
\providecommand \@ifxundefined [1]{%
 \@ifx{#1\undefined}
}%
\providecommand \@ifnum [1]{%
 \ifnum #1\expandafter \@firstoftwo
 \else \expandafter \@secondoftwo
 \fi
}%
\providecommand \@ifx [1]{%
 \ifx #1\expandafter \@firstoftwo
 \else \expandafter \@secondoftwo
 \fi
}%
\providecommand \natexlab [1]{#1}%
\providecommand \enquote  [1]{``#1''}%
\providecommand \bibnamefont  [1]{#1}%
\providecommand \bibfnamefont [1]{#1}%
\providecommand \citenamefont [1]{#1}%
\providecommand \href@noop [0]{\@secondoftwo}%
\providecommand \href [0]{\begingroup \@sanitize@url \@href}%
\providecommand \@href[1]{\@@startlink{#1}\@@href}%
\providecommand \@@href[1]{\endgroup#1\@@endlink}%
\providecommand \@sanitize@url [0]{\catcode `\\12\catcode `\$12\catcode
  `\&12\catcode `\#12\catcode `\^12\catcode `\_12\catcode `\%12\relax}%
\providecommand \@@startlink[1]{}%
\providecommand \@@endlink[0]{}%
\providecommand \url  [0]{\begingroup\@sanitize@url \@url }%
\providecommand \@url [1]{\endgroup\@href {#1}{\urlprefix }}%
\providecommand \urlprefix  [0]{URL }%
\providecommand \Eprint [0]{\href }%
\providecommand \doibase [0]{http://dx.doi.org/}%
\providecommand \selectlanguage [0]{\@gobble}%
\providecommand \bibinfo  [0]{\@secondoftwo}%
\providecommand \bibfield  [0]{\@secondoftwo}%
\providecommand \translation [1]{[#1]}%
\providecommand \BibitemOpen [0]{}%
\providecommand \bibitemStop [0]{}%
\providecommand \bibitemNoStop [0]{.\EOS\space}%
\providecommand \EOS [0]{\spacefactor3000\relax}%
\providecommand \BibitemShut  [1]{\csname bibitem#1\endcsname}%
\let\auto@bib@innerbib\@empty
%</preamble>
\bibitem [{Note1()}]{Note1}%
  \BibitemOpen
  \bibinfo {note} {Alternatively one could consider only the points $x \ge 0$
  and alter the definition of $\protect \mathaccentV {hat}05E{P}^2$ at the
  first lattice point as done in \cite {KL}}\BibitemShut {NoStop}%
\bibitem [{Note2()}]{Note2}%
  \BibitemOpen
  \bibinfo {note} {Similar difference equations occur in polymer quantum
  mechanics, e.g. (V.3) in \cite {poly} and (43) in \cite {gks}.}\BibitemShut
  {Stop}%
\bibitem [{Note3()}]{Note3}%
  \BibitemOpen
  \bibinfo {note} {A brief description of the continued fraction method is the
  following. One may write the difference equation (\ref {eq:recurs1}) as
  \begin {equation}F_k C_k-C_{k-1} = C_{k+1}, \end {equation}where $F_k \equiv
  2-\mu ^2 \Delta - f_k$. Then defining $\alpha _k = C_{k+1}/C_k$ we have the
  recursion relation \begin {equation}\alpha _{k-1} = (F_k - \alpha _k)^{-1}.
  \end {equation}Now, at large $k$ where the asymptotic solution (\ref
  {largek}) holds, we have $\alpha _k \approx A$. Iterating downward from here,
  one can check whether: $F_1 = \alpha _1$ for the boundary condition $C_0 =
  0$; $F_1-1 = \alpha _1$ for the boundary condition $C_0 =
  C_{-1}$.}\BibitemShut {Stop}%
\bibitem [{Note4()}]{Note4}%
  \BibitemOpen
  \bibinfo {note} {The authors are grateful to Andrei Frolov for this
  suggestion.}\BibitemShut {Stop}%
\bibitem [{Note5()}]{Note5}%
  \BibitemOpen
  \bibinfo {note} {Choosing a lattice with $\sigma > 0$ has a similar
  effect.}\BibitemShut {Stop}%
\end{thebibliography}%


\begin{thebibliography}{mybib}

\bibitem{HKK} P. H\'aj\'i\v{c}ek, B. Kay and K. Kucha\v{r}, ``Quantum collapse of a self-gravitating shell: Equivalence to Coulomb scattering,'' Phys. Rev. D {\bf 46} 5439 (1992).

\bibitem{Hajicek} P. Hajicek, ''Quantum mechanics of gravitational collapse,'' Comm. Math. Phys. {\bf 150} 545 (1992).

\bibitem{Berezin} V. Berezin, ''Quantum Black Hole Model and Hawking's Radiation,''  Phys. Rev. D {\bf 55} 2139 (1997), arXiv:gr-qc/9602020.

\bibitem{Stojkovic} J.E. Wang, E. Greenwood and D. Stojkovic, ''Schr\"odinger formalism, black hole horizons, and singularity behavior,'' Phys. Rev. D {\bf 80} 124027 (2009), arXiv:0906.3250 [hep-th].

%\bibitem{Krauss}T. Vahaspati, D. Stojkovic and L. Krauss, ''Observation of Incipient Black Holes and the Information Loss Problem'', Phys.Rev.D76:024005,2007 [gr-qc/0609024]

\bibitem{Vachaspati} T. Vahaspati, D. Stojkovic, ''Quantum Radiation from Quantum Gravitational Collapse,'' Phys. Lett. B {\bf 663} 107 (2008), arXiv:gr-qc/0701096;
T. Vachaspati, ''Schr\"odinger Picture of Quantum Gravitational Collapse,'' Class. Quant. Grav. {\bf 26} 215007 (2009), arXiv:0711.0006 [gr-qc].

\bibitem{Xu} D. Xu, ``Quantum collapse of a self-gravitating thin shell and statistical model of quantum black hole,''
        Physics Letters B {\bf 641} 221 (2006).

\bibitem{poly} A. Ashtekar, S. Fairhurst and J.L. Willis,
Class. Quant. Grav. {\bf 20} 1031 (2003), arXiv:gr-qc/0207106.

\bibitem{eric}  E. Poisson, {\it A Relativist's Toolkit}, Cambridge University Press, Cambridge (2004).

\bibitem{HL} V. Husain and J. Louko, ``Quantum gravity and the Coulomb potential,'' Phys. Rev. D  {\bf 76} 084002 (2007), arXiv:0707.0273 [gr-qc].

\bibitem{KL} G. Kunstatter and J. Louko, ``Boundary conditions in quantum mechanics on the discretized half-line,'' J. Phys. A: Math. Theor. {\bf 45} 305302 (2012), 	arXiv:1201.2886 [gr-qc].

\bibitem{gks} J. Gegenberg, G. Kunstatter and R.D. Small,``Quantum Structure of Space Near a Black Hole Horizon,''
 Class. Quant. Grav. {\bf 23} 6087 (2006), arXiv:gr-qc/0606002.
 
\bibitem{CFL} R. Courant, K. Friedrichs and H. Lewy, ``On the Partial Difference Equations of Mathematical Physics,'' IBM J. \textbf{11} 215 (1967).

\bibitem{israel} W. Israel, ``Singular Hypersurfaces and Thin Shells in General Relativity'', Nuovo Cimento {\bf 44B} 1 (1966).

\end{thebibliography}
\end{document}